\documentclass[aps,pre,twocolumn,groupedaddress,superscriptaddress,showpacs]{revtex4-1}
\usepackage{graphicx,multirow,amsmath,url,txfonts}

\begin{document}

\title{Collective decision making with a mix of majority and minority seekers}

\author{Petter Holme}
\affiliation{Department of Energy Science, Sungkyunkwan University, Suwon 440--746, Republic of Korea}
\author{Hang-Hyun Jo}
\affiliation{BK21plus Physics Division and Department of Physics, Pohang University of Science and Technology, Pohang 37673, Republic of Korea}
\affiliation{Department of Computer Science, Aalto University School of Science, P.O. Box 15400, Espoo, Finland}

\begin{abstract}
We study a model of a population making a binary decision based on information spreading within the population, which is fully connected or covering a square grid. We assume that a fraction of the population wants to make the choice of the minority, while the rest want to make the majority choice. This resembles opinion spreading with ``contrarian'' agents, but has the game theoretic aspect that agents try to optimize their own situation in ways that are incompatible with the common good. When this fraction is less than $1/2$, the population can efficiently self-organize to a state where agents get what they want---the majority (i.e.\ the majority seekers) have one opinion, the minority seekers have the other. If the fraction is larger than $1/2$, there is a frustration in the population that dramatically changes the dynamics. In this region, the population converges, through some distinct phases, to a state of approximately equal-sized opinions. Just over the threshold the state of the population is furthest from the collectively optimal solution.
\end{abstract}


\maketitle

\section{Introduction}

Simple models of human behavior can help us understand population-wide phenomena in society~\cite{castellano}. The purpose of such models is usually not to forecast some social process, but rather to link behavioral causes to population-level effects, in other words, to explore possible social mechanisms. In this paper, we combine two well-studied scenarios---collective decision making and the minority game. We assume that agents are faced with a binary decision to choose one of two ``opinions''. We assume that they can access the current opinion of other agents. When the information is consistent enough, they decide on one opinion and never revert it. This scenario of decision making was also studied in Ref.~\cite{gronlund}. We add another element to this scenario, namely that a fraction $\phi$ of the agents that would prefer minority choice~\cite{challet}. In the minority game, all agents want to be in the minority. Having a mix of minority and majority seekers introduces an interesting tension in the decision making~\cite{de_martino}. In statistical physics terms, the system becomes frustrated---the constraints on the system force the configuration to be locally suboptimal.

To study this situation, we need additional assumptions about what information is accessible to the agents. We assume that they are unaware of the fraction of minority seekers, and we argue that for a majority seeker a good and simple strategy is to decide on the opinion that it perceives is prevailing in the population, while minority seekers decide on the opposite opinion. This idea has been explored for the voter model~\cite{sudo,masuda}, where it has been shown to fundamentally change the model behavior. As we will see, this strategy fares better, on average, than other simple strategies like choosing a random opinion independent of the others' opinion or just sticking to one opinion.

There are many papers investigating the effects of \textit{contrarians}---agents that somehow act in the exact opposite way to the others. In the voter model, this behavior is implemented in a similar way to our model~\cite{sudo,masuda}. The difference is that the agents in a voter model only remember the last interaction, while in our model they are able to integrate the information for an arbitrarily distant past. The same is true for the majority voter model~\cite{liggett}, where an agent updates its opinion to the majority of its neighborhood. In this case, contrarian agents can change their opinion to a minority opinion if they have a higher number of neighbors, i.e., degree, than the average~\cite{li}. Another model of opinion dynamics with contrarian agents was proposed by Galam~\cite{galam2}. In that model, the opinions are spread by normal agents taking the majority opinion of randomly sampled groups, and the contrarians taking the minority opinion. In Ref.~\cite{galam2}, Galam argues that this can explain surprising voting results. Our problem differs from this model in three aspects. First, our agents can remember more information of their past encounters. Second, our model could straightforwardly be applied to arbitrary interaction topologies, like networks. Third, our agents are selfish in the vein of game theory (in particular, the minority game~\cite{challet}). Indeed, the minority with contrarian agents have been studied~\cite{zhong}, which is however a quite different problem in that the contrarian agents differ in their strategies, not their goals.

In this paper, we will first derive the model, then get some ideas about the dynamics from examples, and finally study the results for all parameter values, except the number $N$ of agents. Indeed, all parameters, including $N$, would be interesting to study, but to restrict the scope we make a case study for a large value of $N$, and leave the asymptotics for future studies.

\begin{figure*}
\includegraphics[width=\linewidth]{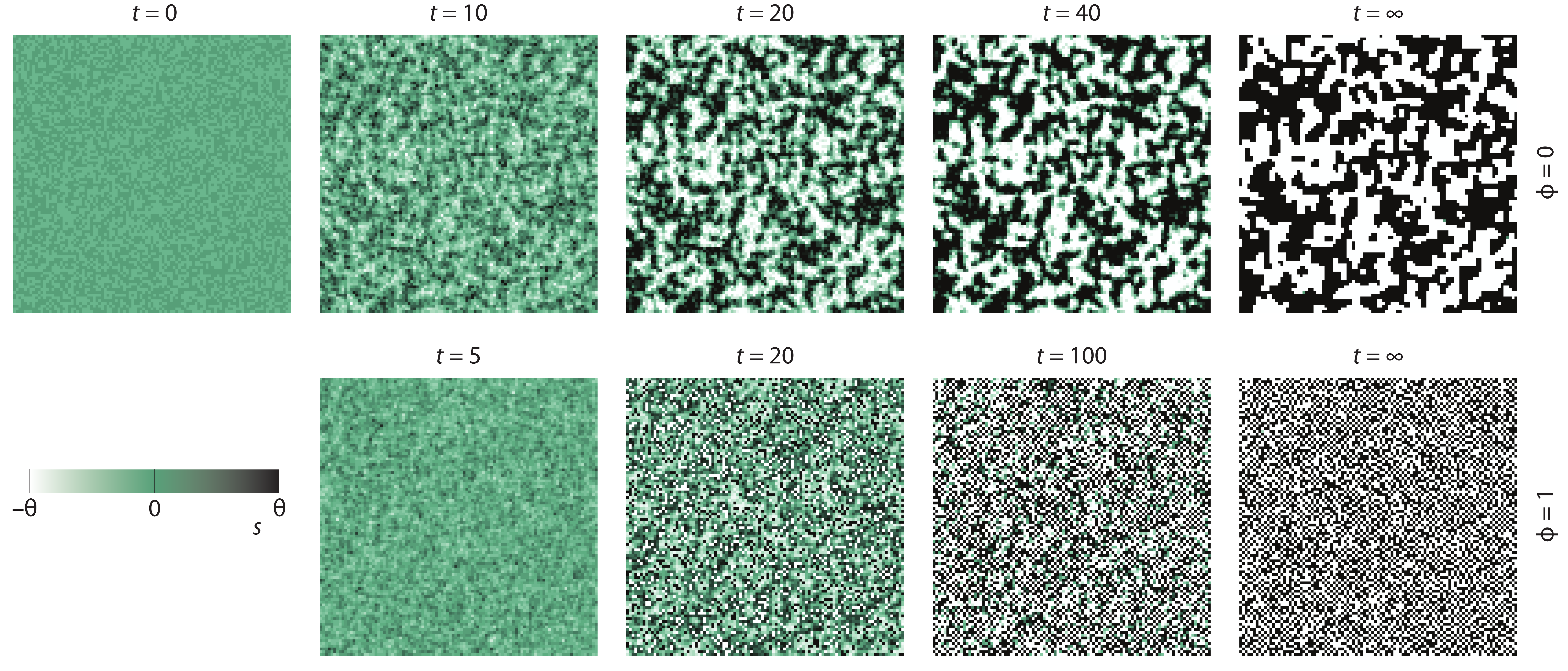}
\caption{Illustration of the behavior of the spatial version of the model for the extreme values of $\phi$. The upper row shows the time evolution of the state variable $s_i$ for a situation when all agents are majority seekers. The lower row shows the situation with only minority seekers. The initial condition is $p=1/2$, i.e.\ the $s_i$ is set to $1$ or $-1$ with equal probability. The decision threshold $\theta$ is $10$.}
\label{fig:ill}
\end{figure*}

\section{Deriving the model}

We assume that $N$ agents are faced with a binary decision. Like in an election, once they have made their decision they cannot undo it. The agents gather information about the decision from interaction with others, either on a complete graph or on a two-dimensional lattice. Once they think the information is conclusive enough, they finalize their decision. In the model, we represent the current state of gathered information by an integer state variable $s_i$ for each agent $i$. The initial configuration is a random distribution of $\pm1$, i.e., $1$ with probability $p$, otherwise $-1$. $p$ can thus control the initial information content, or the general sentiment in the population when the decision process starts. The magnitude of $s_i$ represents the consistency of the information that the agent $i$ has been exposed to. If the magnitude of $s_i$ exceeds an integer threshold value $\theta$, which we call a decision threshold, the agent makes the decision (whereupon it is never changed). During an interaction, an agent influences the other by the sign of $s_i$, where we assume that the magnitude of $s_i$ does not matter. If $i$ interacts with $j$ at time $t$, then
\begin{equation}\label{eq:update_s}
s_i(t+1) = s_i(t) + \mathrm{sgn} \, s_j(t)
\end{equation}
where
\begin{equation}
\mathrm{sgn} \, x = \left\{ \begin{array}{rl} 1 & \mbox{if $x>0$}\\0 & \mbox{if $x=0$} \\ -1 & \mbox{if $x<0$} \\ \end{array}\right..
\end{equation}
In~\cite{gronlund}, Gr\"onrund~\textit{et al.} studied this model on various network topologies when all agents are majority seekers, i.e., $\phi=0$. They showed that the population can efficiently integrate the initial information. To implement the minority or majority seeking, we note that $\phi=1/2$ divides the parameter space into two game-theoretically different regions. If $\phi<1/2$ (i.e.\ a majority of the population is majority seekers), then a minority seeker can just reverse its opinion when it makes the decision. In other words, if $|s_i|=\theta$ during the updates by Eq.~(\ref{eq:update_s}), then $s_i$ is replaced by $-s_i$. If $\phi>1/2$, then the situation becomes much more complex. Without knowledge about which agents have fixed their opinions, it is probably hard to beat a random choice of $s_i$. However, since we also assume that the agents are unaware of $\phi$, we need to have the same strategy throughout the parameter space. Thus we stick to the strategy for all $\phi$-values---also minority seekers reversing their opinion when fixing their opinion (i.e.\ taking a decision). We will later argue that this strategy is more efficient than simpler strategies on average.

\begin{figure*}
\includegraphics[width=0.9\linewidth]{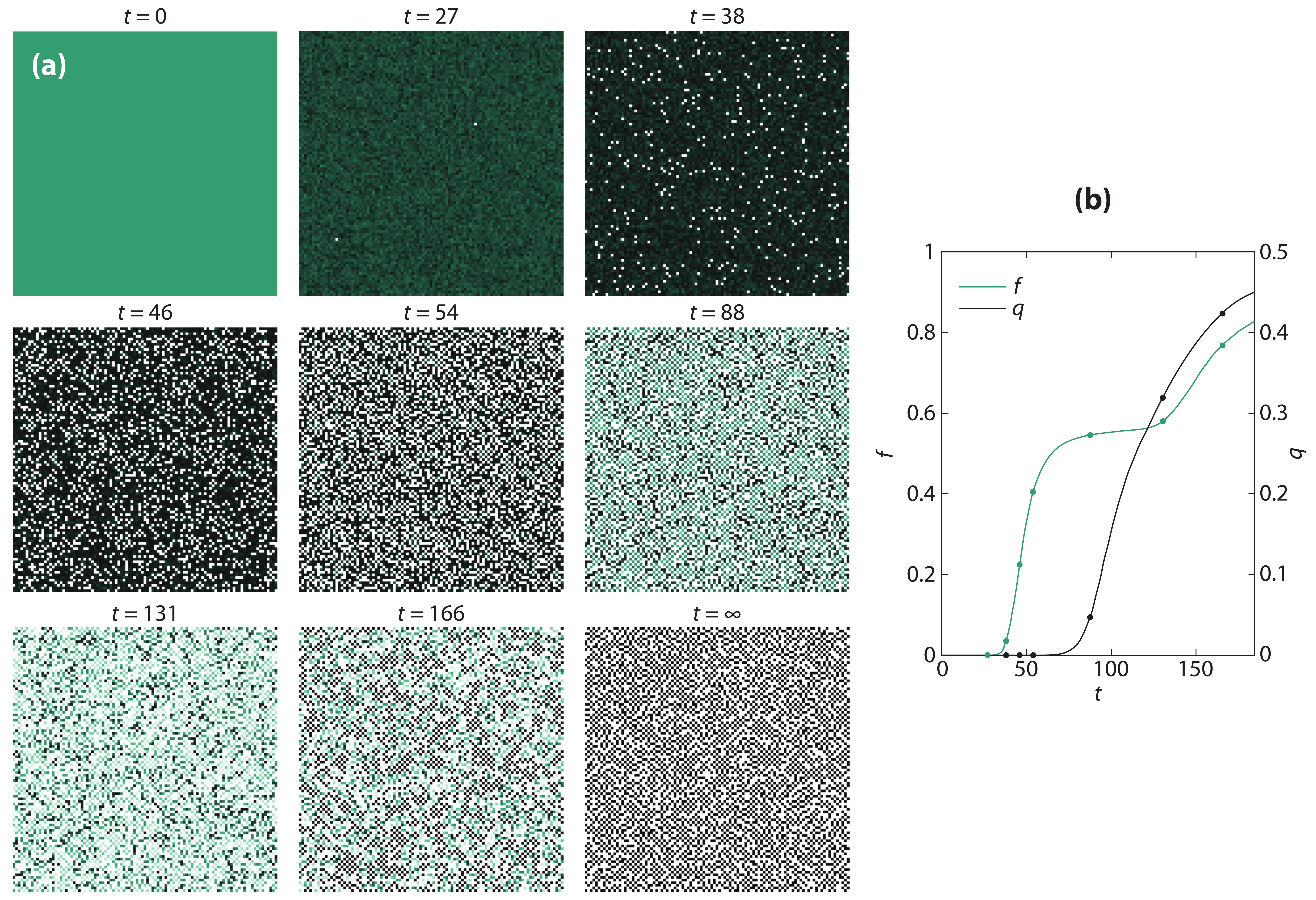}
\caption{Illustration of the all-minority-seeker case with a larger decision threshold ($\theta=50$) which separates the time evolution into different phases. Panel (a) shows snapshots of the population for different times. Panel (b) shows the time evolution of the average opinion $q$ and the fraction of agents with fixated opinions $f$. The dots on the curves mark the snapshots in panel (a). Here we took $p=1$ so that initially all agents have $s_i=1$.}
\label{fig:ill2}
\end{figure*}

We summarize the model as follows:
\begin{enumerate}
\item For all $i$, initialize $s_i$ to $1$ with probability $p$, otherwise $-1$.
\item \label{step:pick} Pick a random agent $i$ and a random neighbor $j$ of $i$.
\item If $|s_i|<\theta$, update $s_i$ according to Eq.~(\ref{eq:update_s}).
\item Increment time $t$ to $t+1/N$ (to let one unit of time represent the average time between updates of $s_i$ for a particular $i$).
\item If now $|s_i|=\theta$, then with a probability $\phi$ replace $s_i$ by $-s_i$.
\item Unless all agents have $|s_i|=\theta$, repeat from step~\ref{step:pick}.
\end{enumerate}
We can think of $s_i$ as performing a random walk on the integers until it reaches a distance $\theta$ where it gets stuck (possibly also changing sign). The probability of going in the positive direction depends on the states of other agents, which makes standard analytically solutions hard. It is straightforward to generalize the model to an arbitrary topology---in step~\ref{step:pick}, just pick $j$ as a neighbor of $i$. In this paper, we will use a fully-connected or well-mixed topology and a square grid with periodic boundary conditions. A NetLogo implementation of the square-grid version of the model can be found at \url{http://petterhol.me/majority-minority-decision-making/}.

We use $N=10,000$ agents, implying a lattice of size $100\times 100$ for square grids, and $10^4$--$10^5$ independent runs for the statistical analysis. These give so good statistics that error bars representing standard error would be smaller than the symbol size in all our plots and are thus omitted.

\section{Numerical results}

We will analyze the model numerically through three quantities: the average opinion $q$, i.e., the average value of $\mathrm{sgn}\, s_i$, the average fraction $\sigma$ of agents in the desired opinion group (majority seekers in the majority, minority seekers in the minority), and the fraction $f$ of agents with $|s_i|=\theta$.

\begin{figure*}
\includegraphics[width=\linewidth]{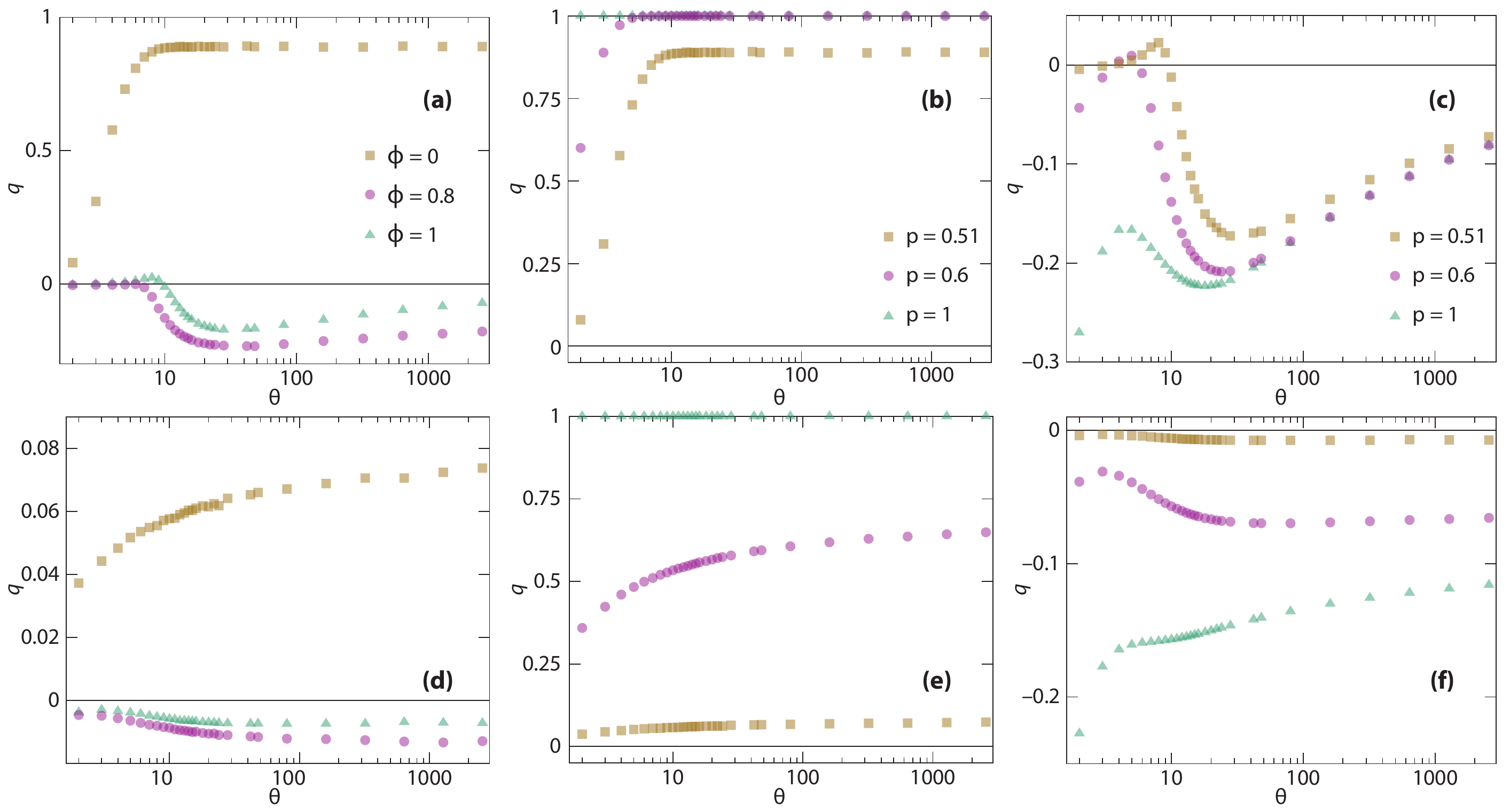}
\caption{The average opinion $q$ depending on the decision threshold $\theta$. Panels (a)--(c) represent fully connected networks; panels (d)--(f) show the results on square grids. Panels (a) and (d) show $q$ for various values of $\phi$ and $p=0.51$; panels (b) and (e) show the results for the population of only majority seekers ($\phi=0$) and various values of $p$; panels (c) and (f) show the results for the population of only minority seekers ($\phi=1$).}
\label{fig:q_vs_theta}
\end{figure*}

\begin{figure*}
\includegraphics[width=\linewidth]{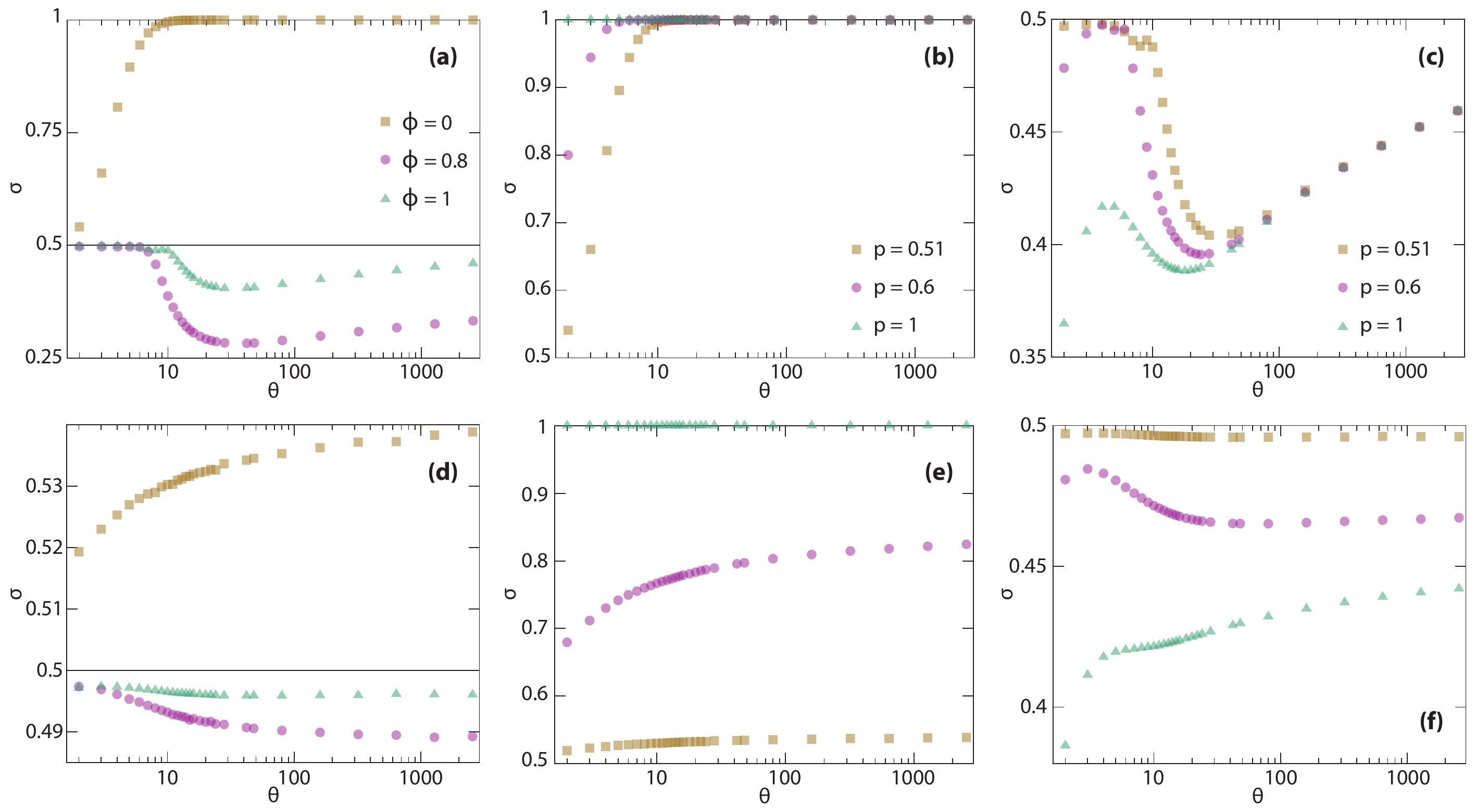}
\caption{The ratio of satisfied agents $\sigma$ as a function of the decision threshold $\theta$. The panels correspond to the panels of Fig.~\ref{fig:q_vs_theta}.}
\label{fig:sat_vs_theta}
\end{figure*}

\subsection{Time evolution}

Our model defines a non-equilibrium process that eventually reaches a fixed state where eventually $s_i=\pm\theta$ for all $i$. This construction is more than just for convenience---since we are modeling a decision process (influenced by opinion spreading) it makes sense that all agents eventually make fixed decisions. Many opinion spreading models, like the voter model~\cite{clifford,liggett,masuda,sudo}, also share this feature, although there is no explicit rule of decision in the original voter model. Maybe it is not so surprising that for some parameter values, the configurations of our model resemble those of the voter model. See Fig.~\ref{fig:ill} for a simulation on a square-lattice topology with $\theta=10$ and $p=1/2$. For $\phi=0$, i.e., when all agents are majority seekers, the model converges to the characteristic coarse-grained patches of the spatial voter model~\cite{clifford}. A larger $\theta$ will give larger patches and smoother edges of the patch boundaries. In the $\theta\rightarrow\infty$ limit the patch boundaries will have zero curvature, i.e., they will be bands across the square grid. When $\phi=1$, i.e., all agents are minority seekers, the final state shows a local checkerboard pattern. (By ``locally'' we mean that there are boundaries, like dislocations in a crystal, where the checkerboard pattern is broken.) In the $\theta\rightarrow\infty$ limit the converged configuration is expected to be a perfect checkerboard pattern.

When there is a majority of minority seekers and the decision threshold is comparatively large, the time evolution becomes segmented into distinct phases. In Fig.~\ref{fig:ill2}, we show such a situation starting from a configuration with all agents having $s_i=1$, i.e., $p=1$ and $\theta=50$. Naturally agents start reaching the decision threshold around $t=50$. Since minority seekers revert their opinions when taking their decision, we can see the fixated agents as white squares on an otherwise dark background (black square representing $s_i=\theta$ and white square representing $s_i=-\theta$) appearing in Fig.~\ref{fig:ill2}(a) at $t=27$ and $t=38$. These fixed agents of negative $s_i$ will give a negative influence to the other agents they are interacting with. In other words, the overall rate of decision slows down until $t\approx 130$ when agents start reaching $s_i=-\theta$ threshold. For the square grid, this is an overdamped oscillation. In other words, we will not see more plateaus by extending Fig.~\ref{fig:ill2}(b)---$f''(t)=0$ at three, but not more than three, $t$ values.

Our model has some similarities to the Ising model~\cite{baxter} and related models of disordered systems~\cite{mezard}. When the interaction is ferromagnetic, the ground state has all spins pointing in the same direction (not unlike, but not identical to when $\phi=0$ and $\theta\rightarrow\infty$). For negative interaction, i.e., the anti-ferromagnetic Ising model, the ground state is a checkerboard pattern like $\phi=1$. There differences too---the structure of the configurations change if one tunes $\phi$, whereas intermediate (non-zero) coupling strengths of the Ising model all have the same ground states.

\begin{figure}
\includegraphics[width=\linewidth]{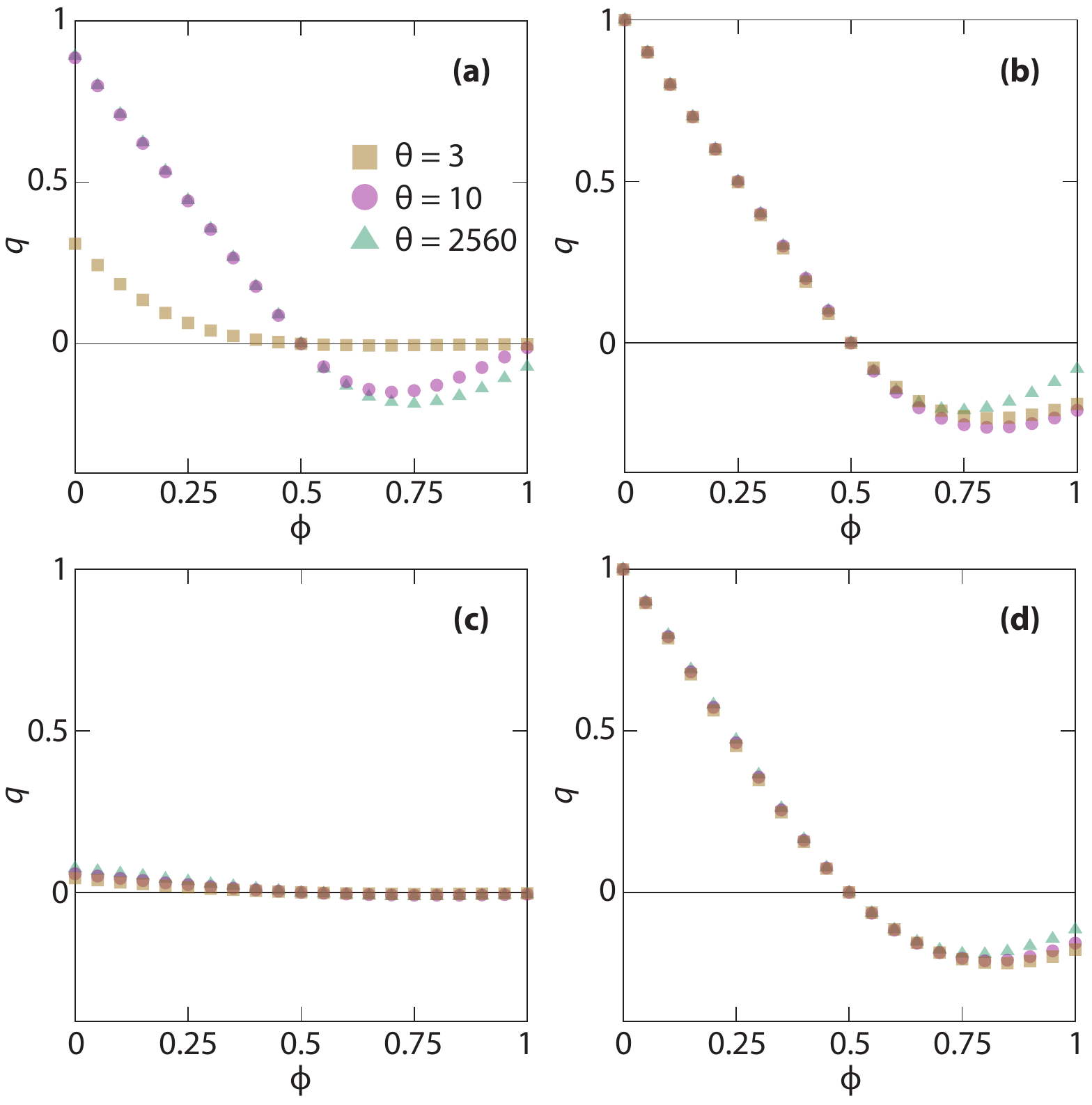}
\caption{The average opinion $q$ as a function of the fraction of minority seekers $\phi$. Panels (a) and (b) show the results for fully connected networks; panels (c) and (d) are the corresponding panels for square grids. Panels (a) and (c) have $p=0.51$; panels (b) and (d) have $p=1$.}
\label{fig:q_vs_phi}
\end{figure}

\begin{figure}
\includegraphics[width=\linewidth]{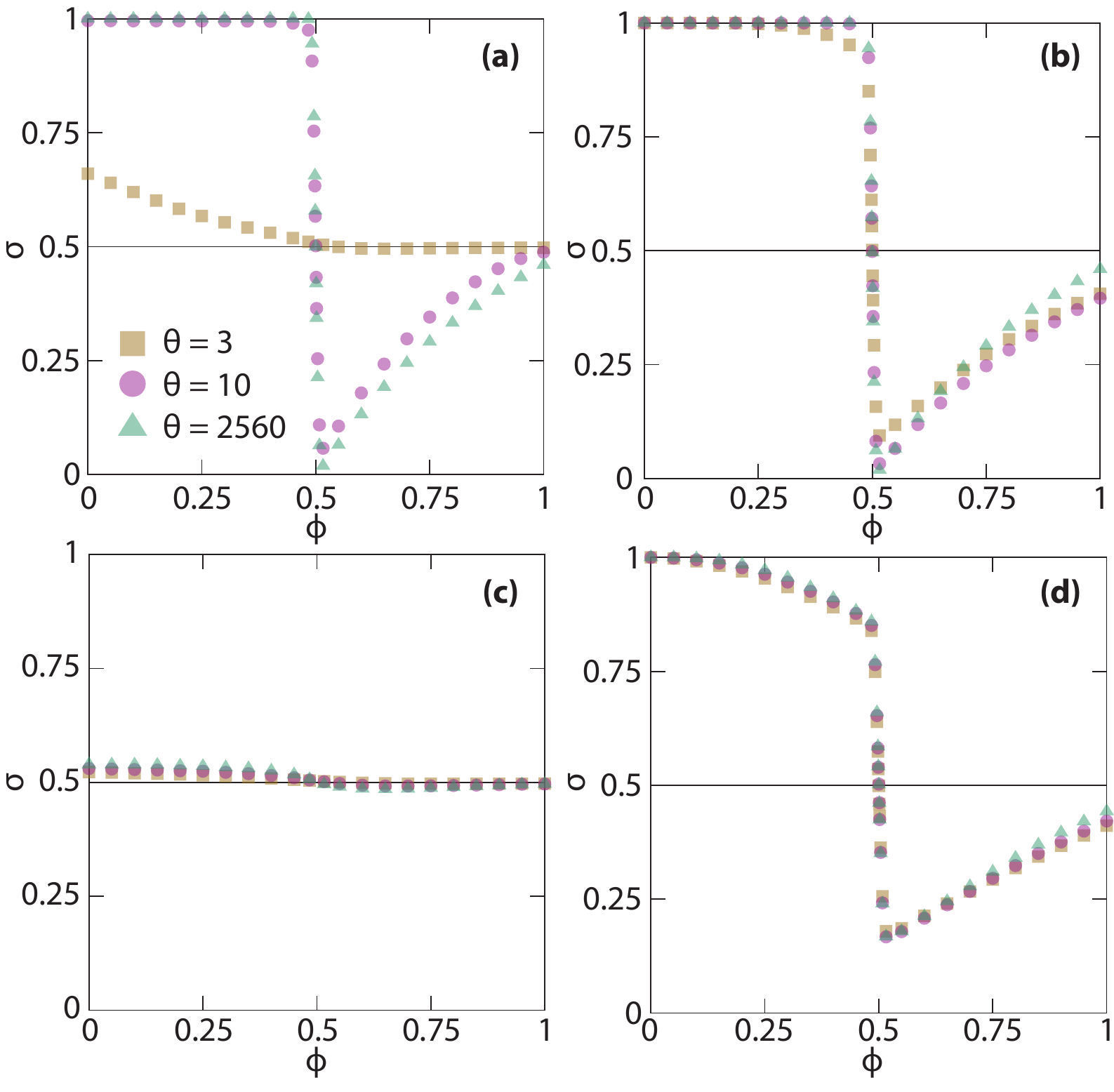}
\caption{Here we investigate how $\sigma$ depends on $\phi$. Panels (a) and (b) display the results for fully connected networks; panels (c) and (d) show the corresponding figures for square grids. (a) and (c) have $p=0.51$; (b) and (d) have $p=1$.}
\label{fig:sat_vs_phi}
\end{figure}

\subsection{Dependence on the decision threshold}

Our first statistical analysis concerns how the average opinion depends on the decision threshold. When $\phi<1/2$, as $\theta$ increases, the average opinion stabilizes fast to a value very close to $q=1-2\phi$, see Fig.~\ref{fig:q_vs_theta}(a). This is true for all $p\gtrsim 0.55$ as can be seen in Fig.~\ref{fig:q_vs_theta}(b) where we plot $q$ for various values of $p$ in a population of only majority seekers. The fact that $q$ levels off at values less than $1$ for $1/2<p\lesssim 0.55$ is probably due to a finite size effect. Even though the binomial distribution of the initial opinion is so narrow that the dynamics of the population extremely rarely starts from a majority of $-1$ for these values, the population could move toward the negative opinion by chance. In other words, even if $p>1/2$, for small enough population sizes, the symmetry is effectively not broken.

When $\phi>1/2$, the situation is dramatically different. Now the average opinion is less than zero for most, but not all, $\theta$ values. First $q$ grows to a maximum around $\theta = 10$ (which is fairly independent of $\phi$), then it drops to a minimum around $\theta = 30$, whereupon $q$ starts increasing towards $0$. This behavior is probably related to the phases seen in of Fig.~\ref{fig:ill2}. How well the information is integrated within the population at the time agents start fixating their opinions depends on $\theta$. If many agents reach the decision threshold $\theta$ around the same time, the minority seekers would choose $\mathrm{sgn}\, s_i=-1$ and the influence of the new negative agents would not be large enough to stop $-1$ from becoming the majority. This can explain the decrease of $q$ to the minimum around $\theta=30$. Another interesting, but hard to explain, observation is that the $q$-curve for $\phi=1$ is not extreme---the lowest $q$-values occur when $\phi\approx 0.8$.

In Fig.~\ref{fig:q_vs_theta}(b) and (c), we look at the extreme cases of only majority or minority seekers, respectively. We see that when $\phi=0$ (only majority seekers, Fig.~\ref{fig:q_vs_theta}(b)) the opinions converge fast with $\theta$. For $\phi=1$, on the other hand, $q(\theta)$ always follows an increase-decrease-increase form. For both $\phi=0$ and $\phi=1$ the curves converge for $p\gtrsim 0.51$ and $\theta\gtrsim 50$. The effect of increasing $\theta$ is obviously to better integrate opinions in the population---the larger $\theta$, the higher ``collective intelligence.'' Apparently this effect expires at some point, presumably all available information is integrated around $\theta=50$.

The lower panels of Fig.~\ref{fig:q_vs_theta} show the results for two-dimensional square grids, corresponding to the upper panels of the same Figure. Most observations for the fully-connected case hold for the square grids too, but the effects are weaker. For example, $q$ does not visually converge as $\theta$ increases at least up to $\theta=2560$. For the all-minority-seekers case, $q$ never becomes positive. 

Another way of characterizing the model is to look at the fraction $\sigma$ of agents that reach their goal, i.e., the fraction of majority seekers to end up in the majority and minority seekers to end up in the minority. The $\sigma$ plots corresponding to Fig.~\ref{fig:q_vs_theta} are shown in Fig.~\ref{fig:sat_vs_theta}. The behavior of $\sigma$ follows the same general picture as $q$. We note that in the fully connected topology, for large enough values of decision threshold and with a majority being majority seekers, all agents will eventually be satisfied. In case with square grids, only large values of $p$ will lead to $\sigma=1$.

It is well known that information spreads slower in square grids---the graph distances scale like $\sqrt{N}$ for square grids compared to the constant for fully connected topology. Many sociophysics models~\cite{castellano} show a qualitatively different behavior on lattices of low-enough dimensions, compared to fully connected topologies. Briefly stated, the weaker coupling at a distance for the low-dimensional lattices makes autocorrelations to decay slower, and thus fluctuations to become larger. In our case, these effects are not so strong that the behavior is fundamentally different for the square grids. The largest difference could be that the minima of $q$ and $\sigma$ as functions of $\theta$ vanish for some values of $p$ (see Figs.~\ref{fig:q_vs_theta} and \ref{fig:sat_vs_theta}).

\begin{figure}
\includegraphics[width=\linewidth]{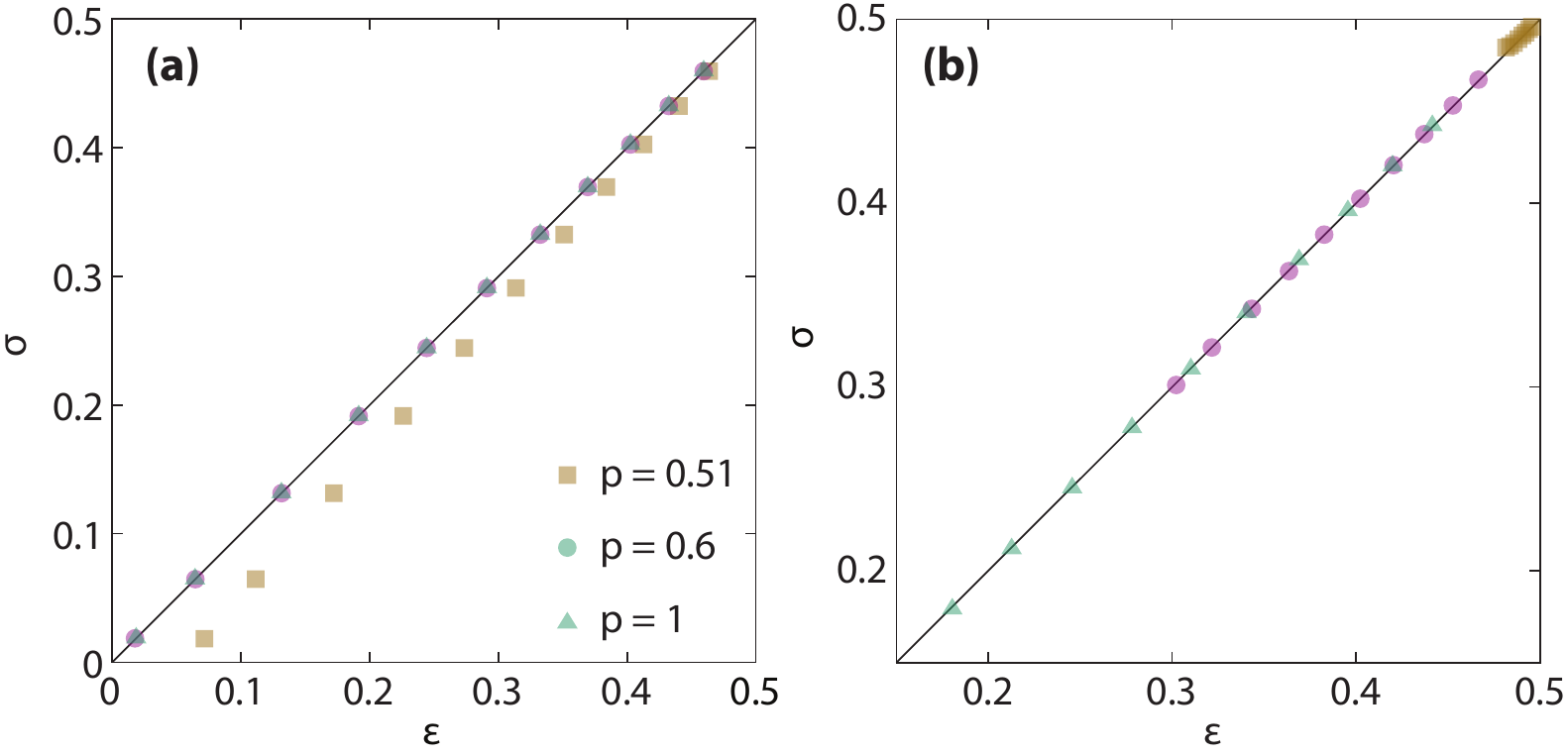}
\caption{Plotting how $\sigma$ depends on $\epsilon$. Panel (a) shows the results for the fully connected topology, and panel (b) displays the results for square grids. $\theta=2560$ was used for both panels. Data points for $\phi\leq 0.505$ are excluded as the theory applies to large $\phi$. The solid line shows $\sigma=\epsilon$.}
\label{fig:sat_vs_eps}
\end{figure}

\subsection{Dependence on the ratio of minority seekers}

We already noted that the dynamics of the model changes dramatically with $\phi$. In particular, $\phi=1/2$ seems to be a boundary between two radically different behaviors. In Fig.~\ref{fig:q_vs_phi}, we plot the average opinion as a function of the fraction of minority seekers. We see that for $\phi>1/2$, $q$ is always negative (i.e., always opposite of the initial average opinion). We also note that the average $q$ over all $\phi$ is positive, meaning that the strategy built into the model is better than trivial strategies like everyone making a random choice, or everyone choosing $s_i=1$. On the other hand, if the agents were aware of $\phi$, a better strategy would be to make a random choice if $\phi>1/2$, and the same as above otherwise.

Comparing the fully connected topology and the square grid, we can once more conclude that the spreading of the information is much slower in the square grid---even for the largest values of $\theta$, if $\phi$ is small, $q$ is still not far from $0$. On the other hand, for the fully connected topology, the population can integrate the information so that $q$ gets close to one.

The behavior of $q$ as a function of $\phi$ is, however, far from as dramatic as the behavior of $\sigma$ (see Fig.~\ref{fig:sat_vs_phi}). For the fully connected case, when $\phi$ crosses $1/2$, $\sigma$ drops from $1$ to $0$ over a small interval. That $\sigma$ can be so low is quite remarkable. It means that almost all majority seekers end up in the minority and vice versa. Slightly below $\phi=1/2$, almost all the majority seekers choose $\mathrm{sgn}\, s_i=+1$, which is the majority opinion, and almost all the minority seekers choose $\mathrm{sgn}\, s_i=-1$. Slightly above $\phi=1/2$, the choice of $s_i$ are almost the same, while the majority opinion is now negative.

For yet larger $\phi$, there is a mix of $\mathrm{sgn}\, s_i=+1$ and $-1$, so that $\sigma$ gets closer to $1/2$. 
In general, since for $\phi>1/2$ the number of agents that cannot be in their desired subpopulation grows linearly with $\phi$, the upper bound of $\sigma$ (if the agents could unselfishly maximize $\sigma$) is
\begin{equation}\label{eq:opt}
\max \sigma = \left\{
\begin{array}{ll}
1 , &  \phi<1/2\\
3/2-\phi ,&  \phi\geq 1/2
\end{array}\right. .
\end{equation}
The effect of the lower dimensionality of the square grid is, as observed above, to dampen the response of parameter changes---this effect is even more dramatic for $\sigma$ than for $q$.

To get a bit more nuanced picture of the dynamics when $\phi$ is large (well above $1/2$), recall that there are two ways an agent can end up with a negative opinion: Either they are minority seekers reaching the positive threshold $\theta$, or they are majority seekers reaching the negative threshold, i.e., $-\theta$. Let $\epsilon$ be the fraction of agents reaching the negative threshold. Then, when the simulations have converged, the states will be distributed as 
\begin{subequations}
\begin{eqnarray}
P(s=\theta)&=& (1-\epsilon)(1-\phi)+\epsilon\phi,\\
P(s=-\theta) &=& (1-\epsilon)\phi+\epsilon(1-\phi),\\
P(s) &=& 0\ \textrm{for}\ -\theta+1\leq s\leq \theta-1,
\end{eqnarray}
\end{subequations}
leading to
\begin{equation}
q=(1-2\phi)(1-2\epsilon),
\end{equation}
or, equivalently (remember, we assume $\phi> 1/2$)
\begin{equation}\label{eq:eps}
\epsilon = \frac{1}{2}-\frac{q}{2-4\phi}.
\end{equation}
Figure~\ref{fig:q_vs_phi} shows that $q$ is always negative for $\phi>1/2$. From Eq.~(\ref{eq:eps}), we can see that this means that $\epsilon<1/2$, i.e., most agents reach the positive threshold. The minority seekers ending up in the minority, contributing to $\sigma$, are those who reach the negative threshold and thus take a positive opinion. The fraction of these agents is $\epsilon\phi$. The majority seekers ending up in majority are those reaching the negative threshold. A fraction $\epsilon(1-\phi)$ belongs to this category. Summing these contributions gives the simple relation
\begin{equation}\label{eq:sigma}
\sigma = \epsilon ,
\end{equation}
which we verify in Fig.~\ref{fig:sat_vs_eps}. It holds very well for all $p$-values larger than $0.55$ and, remarkably, at even lower $p$-values for square grids. Why this approximate calculation works better for square grids is a question we have to leave for future studies.

\section{Discussion}

We have studied a model of collective decision making where a fraction of the population tries to be in the minority. The model shows a complex time evolution and parameter dependence when there is a majority of minority seekers and thus a built-in ``frustration'' in the population by analogy to statistical physics models. When we increase $\phi$ beyond $1/2$, the agents go from a collective optimum where everyone gets what they want, to the worst performance of the model---in some cases almost everyone end up in the group they do not want to be in. One can argue that this is an artifact from the binary nature of the model---in many real situations a minority seeker would not be too disappointed to be in a small majority as in a large majority, so this effect is superficial. On the other hand, there are many ``winner takes it all'' situations in society, voting being one example.

Other simple models of social systems, e.g., in Refs.~\cite{curty} and \cite{galam}, have addressed the coexistence of opinions in collective decision making. Our model gives another mechanism where the goals of the agents with respect to the majority and minority opinions create the coexistence. The special nature of $\phi=1/2$ was also noted in Ref.~\cite{de_martino} who also studied a mixed population of majority and minority seekers, but acting based on previous encounters as in the traditional minority-game setting.

There are interesting future directions from this work. An obvious limitation is the strategy to choose the opinion that creates very low values of the average satisfaction---much lower than the theoretical limit Eq.~(\ref{eq:opt}). Is the theoretical limit unattainable with selfish agents and limited communication? We think so. Note that there is no Nash equilibrium in this model, in the sense that when $\phi>1/2$ there will always be agents that would have fared better if they choose the opposite opinion. Probably selfish agents would never reach over $\sigma=1/2$. At the same time, there might be better, more elaborate strategies, exploiting the information in $|s_i|$, or rather how close to $\theta$ that $|s_i|$ is, to fine-tune the decision. We hope future studies can resolve these questions.

Another issue is how to extrapolate the results to the $N\rightarrow\infty$ limit like in Refs.~\cite{sudo,gronlund}. There are two parameters in our model: $N$ the number of agents which sets the total amount of information in the population and $\theta$ which sets the amount of information one agent can process in the decision process. The most generous assumption, as far as the precision of the decision is concerned, is to assume the limit $N\rightarrow\infty$ and $N/\theta\rightarrow 0$. In this limit, the agents have access to as much information as possible and they also base their decision on as much information processing as possible. In Fig.~\ref{fig:sat_vs_theta}(a), increasing the decision threshold also increases $\sigma$, for the fully connected case. This suggests that the model could reach $\sigma=1$ for $\phi<1/2$ and $\sigma=1/2$ for $\phi>1/2$. For the square grids, such a performance seems unattainable. To establish these limits would be an interesting future direction.

Finally, extending the model beyond binary decisions would be interesting. For some models~\cite{holme_newman}, one needs the number of opinions to be an extensive quantity to see phase transitions. One could also imagine decision making with continuous variables or in higher dimensional spaces.

\begin{acknowledgments}
P.H. was supported by Basic Science Research Program through the National Research Foundation of Korea (NRF) funded by the Ministry of Education (2013R1A1A2011947).
H.-H.J. was supported by Basic Science Research Program through the National Research Foundation of Korea (NRF) funded by the Ministry of Education (2015R1D1A1A01058958).
\end{acknowledgments}

\bibliographystyle{abbrv}
\bibliography{omcof}

\end{document}